# DEVELOPMENT AND TRAINING OF QUANTUM NEURAL NETWORKS, BASED ON THE PRINCIPLES OF GROVER'S ALGORITHM

**Cesar Borisovich Pronin**

**Andrey Vladimirovich Ostroukh**

MOSCOW AUTOMOBILE AND ROAD CONSTRUCTION STATE TECHNICAL UNIVERSITY (MADI)., 64, Leningradsky prospect, Moscow, Russia

**Abstract:** This paper highlights the possibility of creating quantum neural networks that are trained by Grover's Search Algorithm. The purpose of this work is to propose the concept of combining the training process of a neural network, which is performed on the principles of Grover's algorithm, with the functional structure of that neural network, interpreted as a quantum circuit.

As a simple example of a neural network, to showcase the concept, a perceptron with one trainable parameter - the weight of a synapse connected to a hidden neuron.

## Introduction

To illustrate the concept of using Grover's quantum search algorithm for training quantum neural networks, we will use a simple example of a perceptron with one trainable parameter - the weight $w_1$. The goal of the neural network will be to convert the value of the input neuron by multiplying it by the weight $w_1$. The goal of training our neural network will be to determine the correct ratio between the input parameter and the output (based on a specified training dataset), this ratio will be set as the weight $w_1$ of the synapse connecting the input neuron $I_1$ with the hidden $H_1$. In the real world, this task could be implemented, for example, in the form of a dynamic calculator for converting currencies or other values. The given example (Fig. 1) represents a concept that can be scaled up for a task (and a neural network) of any complexity, by adding more inputs and trainable parameters and thus, requiring more quantum bits.

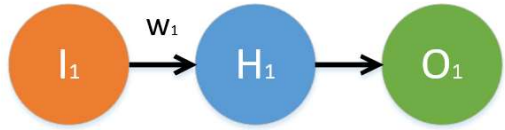

Fig. 1. The proposed example of a neural network.

$$H_{1\ input} = I_1 w_1 \qquad (1)$$

$$H_{1\ output} = f(H_{1\ input}) = H_{1\ input} \qquad (2)$$

The neural network training process will be performed by Grover's algorithm, which will output the weight value $w_1$.

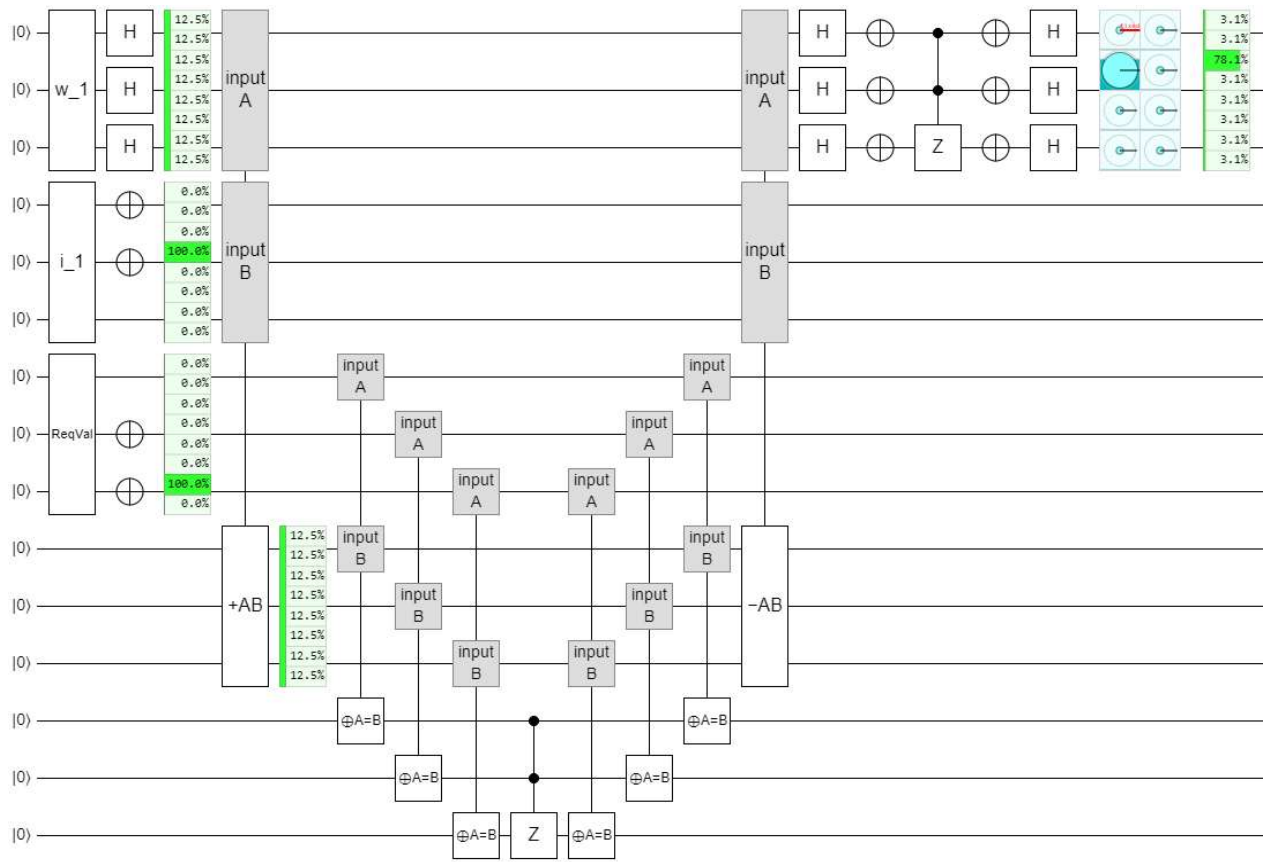

Fig. 2. The quantum neural network training circuit

The proposed quantum perceptron training circuit (Fig. 2) is based on a Grover's algorithm [1-3], in which the structure of the oracle function (function that serves as a search criteria of the Grover's algorithm) is built based on the activation functions of hidden neurons (ex. $F_{act}$ – Fig. 5) and the structure of the neural network itself. The main objective in constructing this training function, is to provide a condition, which could be used in this oracle function to select suitable weight values (ex. comparison operators on Fig. 2 or *Train* - Fig. 5). The input data of the training circuit will consist of prepared training datasets: values of the input neurons $I_1..I_n$, and required output values (in our example: *ReqVal* – Fig. 2). The output data of the algorithm will consist of all possible weight values $w_1..w_n$ that satisfy the training oracle function [4].

The neuron activation function, which in our case is linear and does not contain a certain condition (2), is written inside the oracle of the Grover's algorithm. Additionally, to obtain the training result using Grover's algorithm, a comparison with the required output value was added to the oracle as a condition for choosing suitable values for the weight $w_1$ (Fig. 2).

Instead of carrying out measurement of the results of the training process, we will input the obtained weights in the example neural network (Fig. 1) written in the form of a quantum circuit, and verify the results (Fig. 3). Gates: i_1, w_1, ReqVal, O_1 - are comments that are presented for clarity and do not affect calculations. The input values are coded by sets of Pauli-X operations.

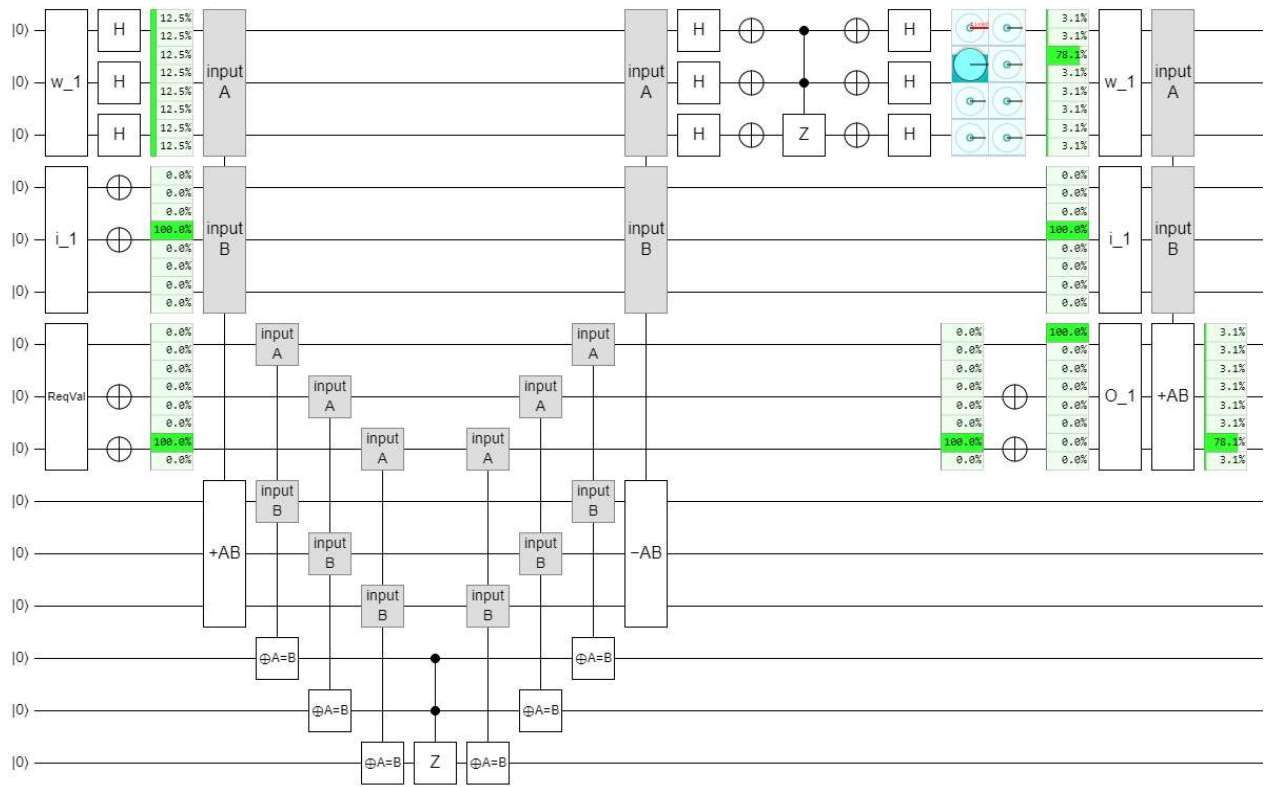

*Fig. 3.a. Quantum neural network – running the training algorithm and verifying the results*[1]

---

[1] Due to binary calculation and comparison operators, it is possible to sometimes receive values that match only within the comparable qubits as a result of the training algorithm. This happens when the multiplication result can not fit in the provided number of qubits, so it is possible to filter out that issue.

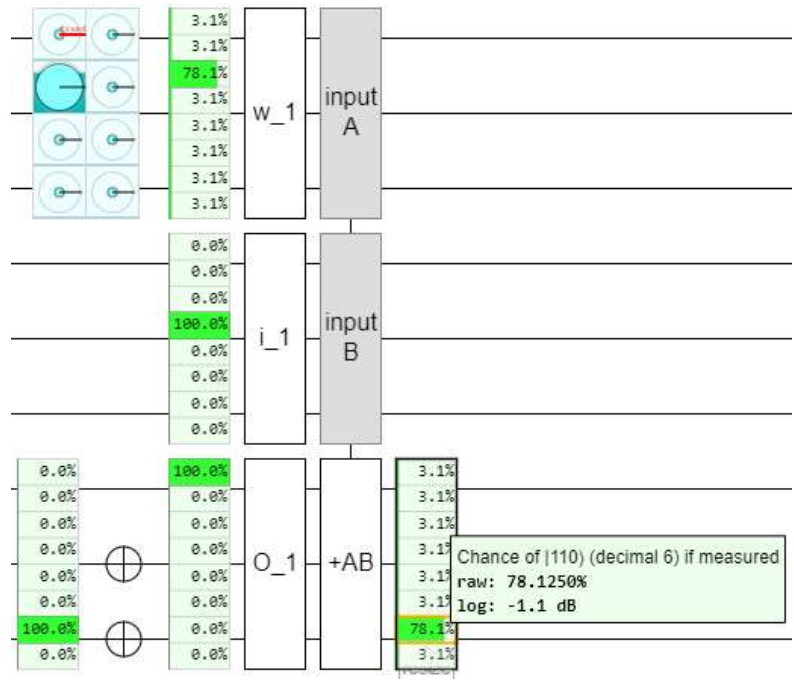

*Fig. 3.b. Quantum neural network – verifying the weight, received as a result of training*

In order to verify training results, it is possible to compare values of output neurons $O_1..O_n$ to the required output values provided by the training datasets $ReqVal_1..ReqVal_n$, by running the neural network with the same input neuron values $I_1..I_n$. For our example (fig. 3), the demonstrated quantum perceptron, built on these concepts, and trained using Grover's algorithm, outputs the value O_1 equal to the required ReqVal for the same input neuron value i_1. The training values were as follows: i_1 = 3, ReqVal = 6, the resulting weight w_1 = 2, the resulting value of the output neuron is: O_1 = 6 = ReqVal.

Therefore, having received the correct weight w_1, it is possible to convert any other values of the input neuron according to the obtained ratio, as in a conventional neural network, after training. For this, the value of i_1, will be changed after passing the training stage of Grover's algorithm, using the reversibility principle. The new input value is i_1 = 5 or $101_2$, and the resulting value is O_1 = 10 or $1010_2$ (Fig. 4).

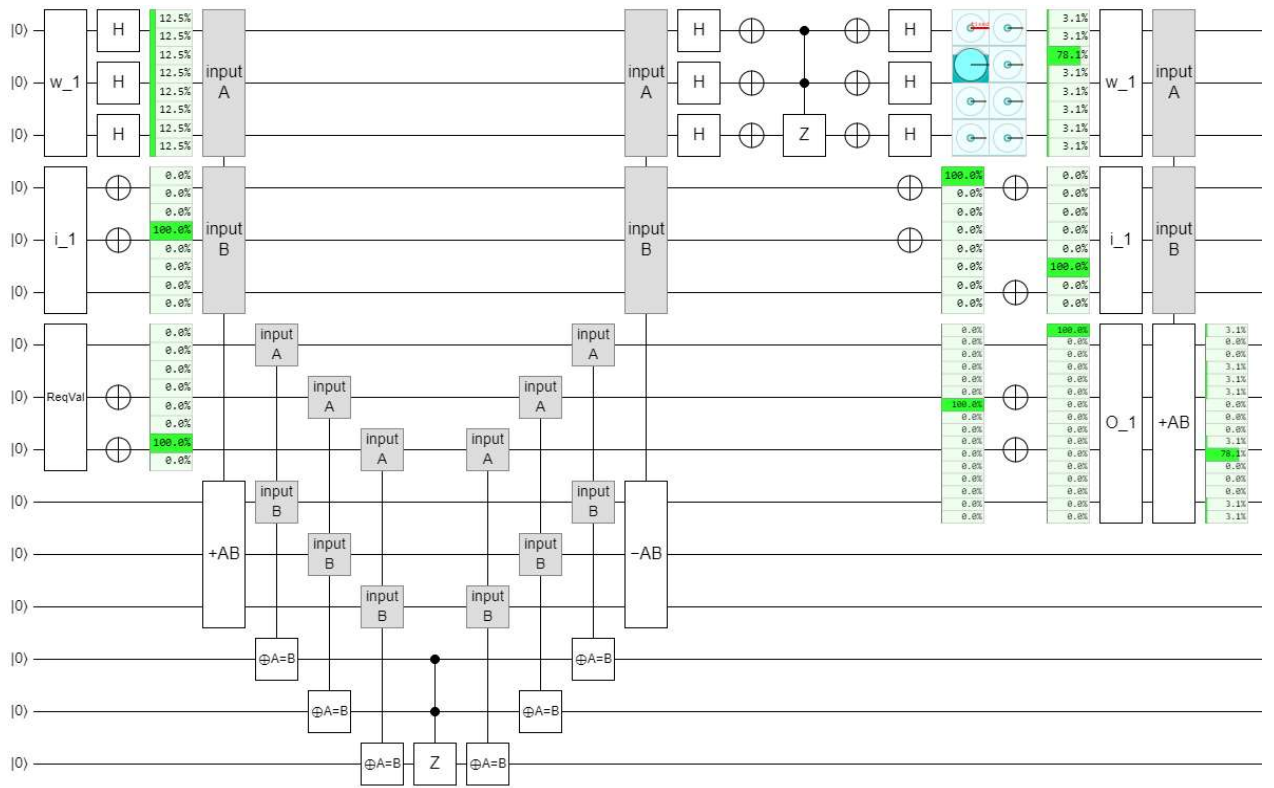

*Fig. 4.a. Quantum neural network – changing the input neuron value after training the neural network.*

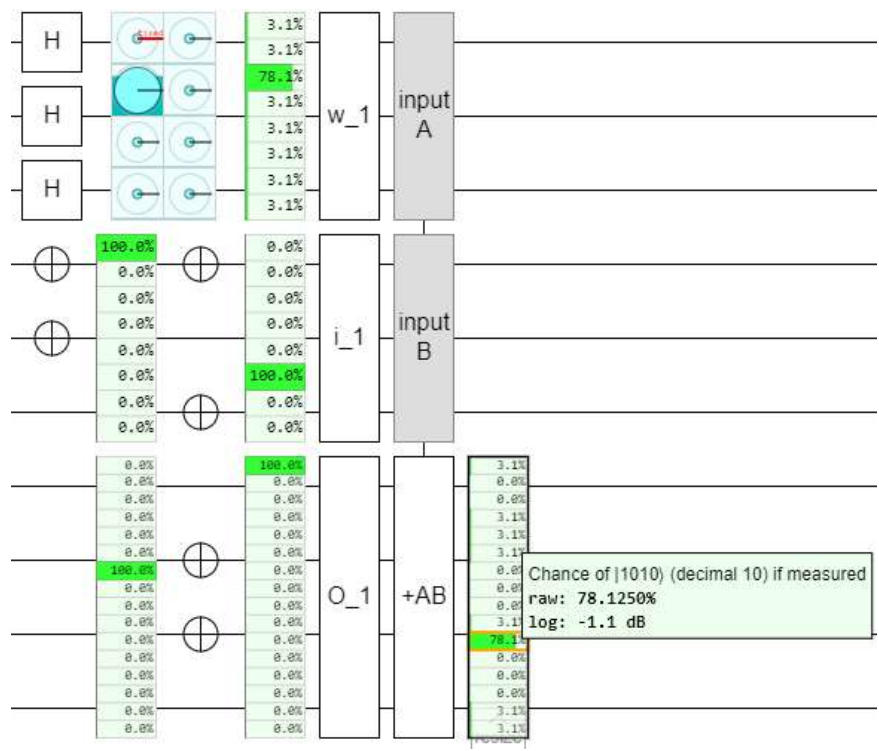

*Fig. 4.b. Quantum neural network – result received by changing the input neuron value after training the neural network.*

Based on these results, it is possible to propose the following generalizing concept for constructing and training quantum neural networks with Grover's algorithm (Fig. 5).

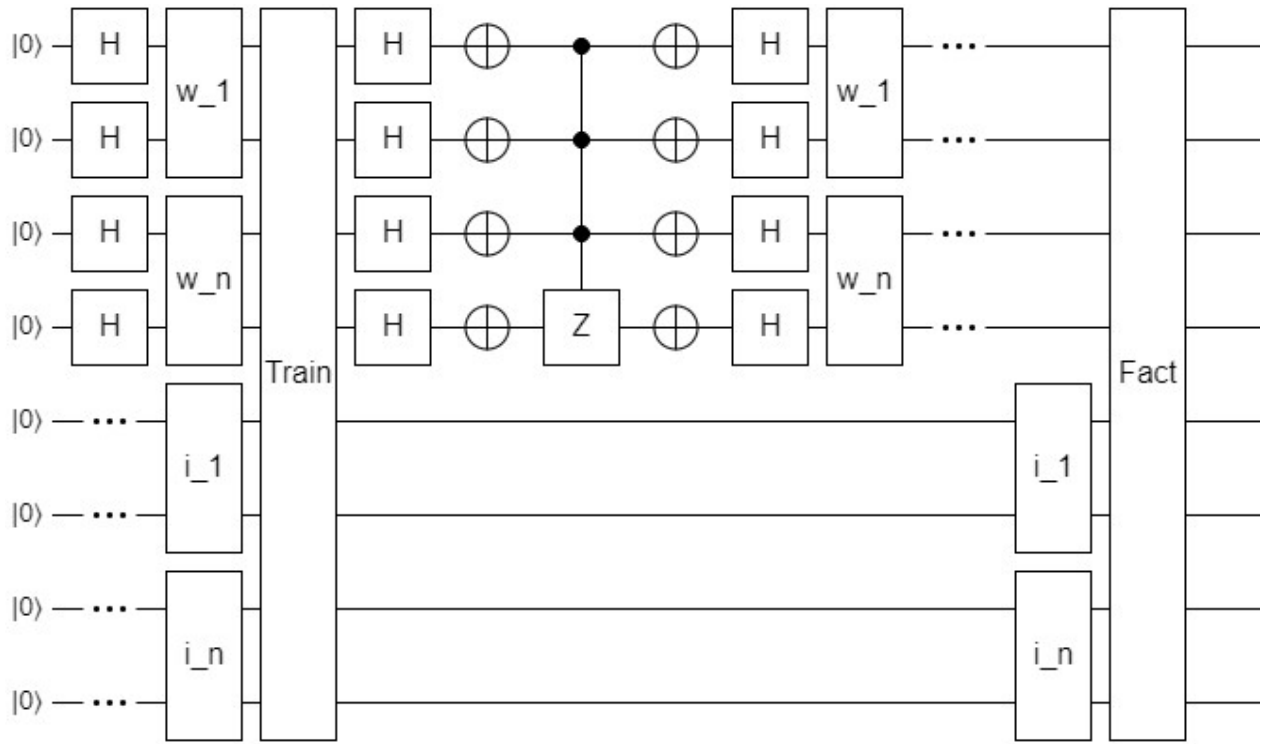

*Fig. 5. The proposed concept of constructing quantum neural networks based on Grover's algorithm.*

## Conclusion

As a result of this research, a method for creating quantum neural networks, trained by utilizing the principles of Grover's algorithm was proposed. The proposed approach can be up scaled to solve larger and more specific problems, requiring access to a larger number of quantum bits. The use of Grover's algorithm in machine learning is a promising direction in the development of quantum computing, since this approach can reduce the number of epochs (iterations of the training process) required to train a neural network.

## References

1) Ostroukh A.V., Pronin C.B.: Researching the Possibilities of Creating Mathematical Oracle Functions for Grover's Quantum Search Algorithm (in Russian) // Industrial Automatic Control Systems and Controllers. - 2018. - No. 9. - P. 3-10. URL: http://asu.tgizd.ru/

2) Cesar Borisovich Pronin, Andrey Vladimirovich Ostroukh: Developing Mathematical Oracle Functions for Grover Quantum Search Algorithm. // Cornell University Library. URL: https://arxiv.org/abs/2109.05921

3) Grover L.K.: A fast quantum mechanical algorithm for database search. // Cornell University Library. URL: https://arxiv.org/abs/quant-ph/9605043

4) Cesar Borisovich Pronin, Andrey Vladimirovich Ostroukh: Development of quantum circuits for perceptron neural network training, based on the principles of Grover's algorithm. // [being processed]

5) Pengcheng Liao, Barry C. Sanders, and Tim Byrnes.: Quadratic Quantum Speedup for Perceptron Training. // Cornell University Library. URL: https://arxiv.org/abs/2109.04695

6) Hiroyuki Tezuka, Kouhei Nakaji, Takahiko Satoh, Naoki Yamamoto: Grover search revisited; application to image pattern matching. // Cornell University Library. URL: https://arxiv.org/abs/2108.10854

7) Quirk // Quirk – online quantum computer simulator. URL: http://algassert.com/quirk

**Author details**

**Andrey Vladimirovich Ostroukh**, Russian Federation, full member RAE, Doctor of Technical Sciences, Professor, Department «Automated Control Systems». State Technical University – MADI, 125319, Russian Federation, Moscow, Leningradsky prospekt, 64. Tel.: +7 (499) 151-64-12. http://www.madi.ru , email: ostroukh@mail.ru , ORCID: https://orcid.org/0000-0002-8887-6132

**Cesar Borisovich Pronin**, Russian Federation, Postgraduate Student, Department «Automated Control Systems». State Technical University – MADI, 125319,

Russian Federation, Moscow, Leningradsky prospekt, 64. Tel.: +7 (499) 151-64-12. http://www.madi.ru , email: caesarpr12@gmail.com , ORCID: https://orcid.org/0000-0002-9994-1032